\documentclass[a4paper,12pt]{article}
\title{Longitudinal resonance and identification of the order
parameter of the A-like phase
 of superfluid $^3$He in aerogel.}
\author {I.A.Fomin\\
P. L. Kapitza Institute for Physical Problems, \\
ul. Kosygina 2, 119334 Moscow,Russia}
\date{ }
\begin{document}
\maketitle
\begin{abstract}
Interpretation of the recent experiments of V.V. Dmitriev et. al.
on longitudinal resonance in the A-like phase, based on specific
properties of "robust" order parameter is proposed.

%\pacs {PACS numbers: 67.57.Lm, 76.50.+g}
\end{abstract}

%\section {Introduction}
In the recent experiments of Dmitriev et al. \cite{dmit1} the
longitudinal NMR in the superfluid  $^3$He in aerogel was first
observed and investigated. The results obtained for the A-like
phase provide new  input for identification of this phase. In
these experiments in parallel with the longitudinal NMR the
transverse line was registered.  For the ensuing discussion the
following results of Ref. \cite{dmit1} are of particular
importance:

1) When the A-like phase is obtained by cooling without specially
applied external perturbations there exist well defined
longitudinal resonance.

 2) Transverse NMR signal at the same conditions is positively
shifted with respect to the Larmor frequency. The line has
two-maximum shape. The shift of one of the two maxima (referred as
$``f"$-line in the Ref. \cite{dmit1}) is approximately 4 times
greater then of the other ($``c"$-line).

3) After application of a series of 180$^o$ pulses on cooling
through T$_c$  both the longitudinal resonance and the $``f"$-line
in the transverse signal disappear. The  intensity of the
$``c"$-line increases.

In what follows it is shown that these qualitative features, in
particular the disappearance of the longitudinal resonance, can be
related to specific properties of the previously proposed "robust"
order parameter \cite{fom1,fom2}:
$$
A^R_{\mu j}=\Delta\frac{1}{\sqrt{3}}e^{i\psi}[\hat d_{\mu}(\hat
m_j+i\hat n_j)+ \hat e_{\mu}\hat l_j].  \eqno(1)
$$
Here $l_j,m_j,n_j$ are mutually orthogonal unit vectors in
momentum space and $d_{\mu}, e_{\mu}$ -- mutually orthogonal unit
vectors in spin space.

NMR probes the form of the order parameter
 via the dipole energy and via the anisotropy
of magnetic susceptibility. For the order parameter Eq. (1) the
dipole energy has the following form:
$$
U_D=\frac{\chi_n}{8 g^2}\Omega^2\left[(\hat{\bf d}\cdot\hat{\bf
m}+\hat{\bf e}\cdot\hat{\bf l})^2+(\hat{\bf d}\cdot\hat{\bf
n})^2+\hat{\bf f}\cdot\hat{\bf n}\right]. \eqno(2)
$$
Here the spin vectors $d_{\mu}, e_{\mu}$ are complemented up to a
triad by introduction of a unit vector $\hat{\bf f}=\hat{\bf
d}\times\hat{\bf e}$.

Dependence of $U_D$ on a relative orientation of $\hat{\bf f}$ and
 $\hat{\bf n}$ reflects the lack of symmetry of the order parameter
 with respect to the interchange of $\hat{\bf d}$ and $\hat{\bf
 e}$.
 Minima of $U_D$ determine possible static orientations of
 the spin triad $\hat{\bf d},\hat{\bf e},\hat{\bf f}$ with respect
 to the orbital $\hat{\bf l},\hat{\bf m},\hat{\bf n}$.
 $U_D$ reaches its absolute minimum when
$$
\hat{\bf d}\cdot\hat{\bf m}+\hat{\bf e}\cdot\hat{\bf l}=0, \quad
\hat{\bf d}\cdot\hat{\bf n}=0, \eqno(3)
$$
$$
\hat{\bf f}\cdot\hat{\bf n}=-1, \eqno(4)
$$
i.e.  $\hat{\bf f}$ is antiparallel to $\hat{\bf n}$. With that
restriction Eqns. (3) determine two mutually degenerate
orientations of $\hat{\bf d},\hat{\bf e}$ in a plane perpendicular
to
 $\hat{\bf n}$. These are $\hat{\bf
d}=\hat{\bf l}$, $\hat{\bf e}=-\hat{\bf m}$ and $\hat{\bf
d}=-\hat{\bf l}$, $\hat{\bf e}=\hat{\bf m}$.

Except for these two minima the dipole energy  has also a
meta-stable minimum
$$
\hat{\bf f}\cdot\hat{\bf n}=1. \eqno(5)
$$
In this case Eqns. (3) are met for arbitrary orientation of
$\hat{\bf d},\hat{\bf e}$ in a plane perpendicular to
 $\hat{\bf n}$, i.e. the minimum is continuously degenerate.
Two configurations: Eq. (4) and Eq. (5) have different NMR
``signatures". For the stable minimum (Ref. \cite{miura}) the
longitudinal resonance frequency is:
$$
\omega^{(s)}_{\parallel}=\Omega    \eqno(6)
$$
and transverse resonance frequency, in a principal order on
$\Omega/\omega_L\ll 1$ is:
$$
\omega^{(s)}_{\perp}=\omega_L+\frac{\Omega^2}{4\omega_L}. \eqno(7)
$$
For the meta-stable orientation because of its continuous
degeneracy
$$
\omega^{(m)}_{\parallel}=0.     \eqno(8)
$$
There is no shift of the transverse resonance frequency from the
Larmor frequency in the order $(\Omega/\omega_L)^2$ :
$$
\omega^{(m)}_{\perp}=\omega_L+O\left[\left(\frac{\Omega}{\omega_L}\right)^4
\right], \eqno(9)
$$
 i.e. the shift is much smaller then for the
stable minimum.

Comparison of these signatures with the results of Ref.
\cite{dmit1} suggests the following qualitative interpretation. In
the experimental cell regions with the stable configuration
$\hat{\bf f}\cdot\hat{\bf n}=-1$ coexist with the regions where
meta-stable configuration $\hat{\bf f}\cdot\hat{\bf n}=1$ is
realized. The
 line $``f"$ in the transverse NMR signal originates
from regions with the stable configuration, while the line $``c"$,
which has much smaller shift -- from the meta-stable. The
longitudinal signal comes from regions with the stable
configuration only. Application of 180$^o$ pulses at cooling
favors formation of the meta-stable configuration. This results in
suppression of the $``f"$-line and in disappearance of the
longitudinal resonance.

The reason why the pulses suppress the stable configuration is not
clear. One can see that application of a 180$^o$ pulse removes a
barrier for conversion of the stable configuration in the
meta-stable. This conversion requires reversal of the direction of
spin ${\bf S}$ with respect to the triad $\hat{\bf d},\hat{\bf
e},\hat{\bf f}$. Motion of the triad $\hat{\bf d},\hat{\bf
e},\hat{\bf f}$ can be parametrized  by Euler angles
$\alpha,\beta,\gamma$ (cf. \cite{fom-rev}), so that $\alpha$ has a
meaning of the phase of precession of ${\bf S}$, $\gamma$ -- the
phase of rotation of the triad $\hat{\bf d},\hat{\bf e},\hat{\bf
f}$ around ${\bf S}$ and $\beta$ - of the tipping angle. Averaging
of the dipole energy over a period of precession results in:
$$
\bar{U}_D=\frac{\chi_n}{8
g^2}\Omega^2\left[(1+\cos\beta)^2\sin^2(\alpha+\gamma)+
\frac{1}{2}\sin^2\beta-\cos\beta\right]. \eqno(10)
$$
The sum $\alpha+\gamma$ is a ``slow variable", it keeps relative
phase of precession and rotation close to the minimum of the
$\bar{U}_D$ for a given tipping angle $\beta$:
$\alpha+\gamma=n\pi$. As a result when $\beta$ is not close to
$\pi$ a relative orientation of ${\bf S}$ with respect to the
triad $\hat{\bf d},\hat{\bf e},\hat{\bf f}$ does not change. At
$\beta=\pi$ the coefficient in front of $\sin^2(\alpha+\gamma)$
turns to zero, angles $\alpha$ and $\gamma$ are decoupled and
during relaxation of ${\bf S}$ to the equilibrium its absolute
value and orientation with respect to the triad $\hat{\bf
d},\hat{\bf e},\hat{\bf f}$ can change. The average dipole energy
for the meta-stable configuration does not depend on
$\alpha+\gamma$ for all $\beta$. There is no symmetry between the
stable and meta-stable configurations with respect to their
reaction to 180$^o$ pulses, but it is not clear why the opposite
conversion is not going on after tipping for 180$^o$. So we have
to take as experimental fact that when 180$^o$ pulses are applied
at cooling the $``c"$-state is formed predominatingly, which
within the suggested identification scheme corresponds to the
meta-stable configuration.

Quantitative comparison reveals certain discrepancies. First, the
shift of the "anomalous" maximum in the transverse resonance data,
being much smaller then the shift of the "normal" maximum, is
still much greater then one would expect from Eq. (9). This
discrepancy could be ascribed to a deviation of the A-like order
parameter realized in particular experiment from the limiting
"robust" form. As an example of possible effect of such deviation
on NMR signal consider the class of axi-planar order parameters,
which are represented in a form:
$$
A_{\mu j}=\Delta e^{i\phi}[\hat d_{\mu}(a_y\hat m_j+ia_z\hat n_j)+
a_x\hat e_{\mu}\hat l_j],  \eqno(11)
$$
where  $a_x, a_y, a_z$ are real numbers restricted by the
normalization condition: $a_x^2+a_y^2+a_z^2=1$. This class
includes the ABM phase ($a^2_y=a^2_z=1/2, a_x=0$) and the "robust"
phase ($a_x^2=a_y^2=a_z^2=1/3$). Guided by the experiment
\cite{dmit1} consider only such deviations from the "robust" form,
which preserve the degeneracy of the meta-stable minimum with
respect to rotations of $\hat{\bf d},\hat{\bf e}$ in the plane
perpendicular to $\hat{\bf n}$. This condition is satisfied if
$a_x=a_y\equiv u, a_z\equiv v$, i.e.
$$
A_{\mu j}=\Delta e^{i\phi}[\hat d_{\mu}(u\hat m_j+iv\hat n_j)+
u\hat e_{\mu}\hat l_j],  \eqno(12)
$$

 The dipole energy for this form of the order parameter is

$$
U_D=\frac{\chi_n}{8 g^2}\Omega^2\left[(\hat{\bf d}\cdot\hat{\bf
m}+\hat{\bf e}\cdot\hat{\bf l})^2+w^2(\hat{\bf d}\cdot\hat{\bf
n})^2+\hat{\bf f}\cdot\hat{\bf n}\right], \eqno(13)
$$
where $w=v/u$. The normalization constant $\Omega^2$ is defined so
that $\omega^{(s)}_{\parallel}=\Omega$. Standard calculation
renders:
$$
\omega^{(m)}_{\parallel}=0  \eqno(14)
$$
and
$$
\omega^{(m)}_{\perp}=\omega_L+(w^2-1)\frac{\Omega^2}{8\omega_L}.
\eqno(15)
$$
For the robust state $w^2=1$ and $\omega^{(m)}_{\perp}=\omega_L$,
but if $w^2>1$ the shift is finite and positive. The shift of the
stable configuration also changes:
$$
\omega^{(s)}_{\perp}=\omega_L+(w^2+1)\frac{\Omega^2}{8\omega_L}.
\eqno(16)
$$

Experimentally measured ratio of the shift of the stable line to
that of the meta-stable can be used for extraction of $w^2$ from
the data:
$$
\frac{\omega^{(m)}_{\perp}-\omega_L}{\omega^{(s)}_{\perp}-\omega_L}=
\frac{w^2-1}{w^2+1}.   \eqno(17)
$$
Assuming the r.h.s. is about 1/4 we arrive at $w^2\approx5/3$.
With that value the order parameter, suitable for description of
the experimental data can be written in a form Eq. (10) with
$u^2=3/11$ and $v^2=5/11$ instead of 1/3 for the "robust" state.

Additional support for the proposed identification comes from the
pulsed NMR data \cite{dmit1}. Dependence of the transverse shift
on the tipping angle $\beta$ for the meta-stable configuration is
given by:
$$
\omega^{(m)}_{\perp}-\omega_L=(w^2-1)\frac{\Omega^2}{8\omega_L}\cos\beta.
\eqno(18)
$$
The $\cos\beta$ dependence agrees with the data obtained in the
experiment \cite{dmit1}. For the stable configuration the tipping
angle dependence is given by:
$$
\omega^{(s)}_{\perp}-\omega_L=\frac{\Omega^2}{8\omega_L}(1+w^2\cos\beta).
\eqno(19)
$$
When $w^2=1$ this formula coincides with that, obtained in Ref.
\cite{miura}. The $1+\cos\beta$ dependence was confirmed in the
experiments with the 97\% aerogel \cite{ishik}. Dependence,
following from Eq. (19) with $w^2=5/3$ also does not contradict to
the data of \cite{ishik}.

Deviation of the order parameter from the robust form switches on
disorienting effect of aerogel, so that the glass-like state could
form.
 This state is based on the nearly robust order parameter
(12) and not on the ABM, as suggested in Ref. \cite{volovik}.  For
the nearly robust order parameter (12) disorienting effect of
aerogel is described by a term proportional to
$(v^2-u^2)\eta_{jl}({\bf r})n_jn_l$, where $\eta_{jl}({\bf r})$ is
a real symmetric traceless random tensor. The disorienting effect
is weakened by a small factor $(v^2-u^2)$. Characteristic length
$L$ of the disordered state is inversely proportional to a square
of this combination, so it must be much greater then in the ABM
phase favoring a situation when $L$ is much greater then the
dipole length and orientation of spin triad $\hat{\bf d},\hat{\bf
e},\hat{\bf f}$ with respect to the orbital $\hat{\bf l},\hat{\bf
m},\hat{\bf n}$ follows a minimum of the dipole energy.
 This property is crucial for the very existence of the
longitudinal resonance. It means also that in a magnetic field
with $\omega_L\gg\Omega$ the orbital vector $\textbf{n}$ in
equilibrium is aligned or counter aligned with the field. Since
$\textbf{n}$ is fixed the disorienting effect of the random
anisotropy is suppressed and disordered state is not formed.

There remains a discrepancy between the value of the dipole
frequency $\Omega$ extracted from the magnitude of the shift of
$``f"$-line  and of the directly measured longitudinal resonance
frequency. According to the data \cite{dmit1} for 29.3 bar at
T=0.835$T_{ca}$ the measured longitudinal resonance frequency
$\omega_{\parallel}$ is only 0.7 of the value of $\Omega$ obtained
from the shift of the $``f"$-line for the same conditions via
Eq.(16) with $w^2=5/3$.  This discrepancy could partly originate
from a difference in definitions of the resonance frequency in Eq.
(6) and in the measurement procedure. Usually the resonance
frequency $\omega_r$ is registered as a maximum of the imaginary
part of magnetic susceptibility $\chi''(\omega,T)$ with respect to
$\omega$ at constant $T$. Formally $\omega_r=\omega_r(T)$ is a
root of equation:
$$
\left(\frac{\partial\chi''(\omega,T)}{\partial\omega}\right)_T=0.
\eqno(20)
$$
In the experiment \cite{dmit1} $\chi''(\omega,T)$ was measured as
a function of $T$ at constant $\omega=\omega_0$. Resonance
temperature $T_r$ was determined as a maximum of
$\chi''(\omega_0,T)$ with respect to $T$:
$$
\left(\frac{\partial\chi''(\omega,T)}{\partial
T}\right)_{\omega=\omega_0}=0 \eqno(21)
$$
and $\omega_0$ was taken as a resonance frequency at $T=T_r$. Two
definitions need not give the same frequency:
$\omega_r(T=T_r)\neq\omega_0$. When the difference
$\nu=\omega_0-\omega_r$ is small it can be expressed in terms of
derivatives of $\chi''(\omega,T)$, taken at $\omega=\omega_r(T)$:
$$
\nu=\frac{(\partial\chi''/\partial T)}{(d\omega_r/d
T)(\partial^2\chi''/\partial\omega^2)}.     \eqno(22)
$$
For the present situation $\nu$ is negative i.e. the resonance
frequency $\omega_r$ and hence $\omega^{(s)}_{\parallel}$  is
greater then its apparent value  $\omega_0$. Crude estimation of
the derivatives renders $\nu\sim\Gamma^2/\omega_r$, where $\Gamma$
is a width of resonance. When $\Gamma\ll\omega_r$ this correction
is negligible, but it is not the case for the longitudinal NMR
line in the A-like phase. Relative correction $\nu/\omega_r$ for
the line, measured in Ref. \cite{dmit1} could be of the order of
0.1, which is a substantial decrease of the discrepancy.

In conclusion, it should be pointed out that the proposed
identification of the observed NMR lines follows naturally from
the assumption of the robust form for the order parameter of the
A-like phase. It gives consistent qualitative description of the
data. Quantitative discrepancies are partly removed by account of
possible deviations of the order parameter from the robust form
and by introduction of the discussed correction in the analysis of
the data. Further NMR experiments, as suggested in Ref.
\cite{dmit1}, will help to make definitive conclusion about the
form of the order parameter of the A-like phase.

I thank V.V. Dmitriev for useful discussions and comments and for
communication of the results of the experiments Ref.\cite{dmit1}
prior to publication. This work is partly supported by RFBR (grant
04-02-16417), Ministry of Science and Education of the Russian
Federation and CRDF (grant RUP1-2632-MO04).

\end{document}